\documentclass[aps,pra,reprint,twocolumn,groupedaddress]{revtex4-1}
\usepackage{times}
\usepackage[normalem]{ulem}
\usepackage{graphicx}
\usepackage{array}
\usepackage{amsmath}
\usepackage{amssymb}
\usepackage{mathrsfs}
\usepackage{color}
\usepackage{soul,xcolor}
\setstcolor{red}

\begin{document}
\title{Correlation Redistribution by Causal Horizons}

\author{L. Pipolo de Gioia}
\affiliation{Instituto de F\'\i sica ``Gleb Wataghin'', Universidade Estadual de Campinas, Campinas, SP, Brazil}
\author{M. C. de Oliveira }\
\affiliation{Instituto de F\'\i sica ``Gleb Wataghin'', Universidade Estadual de Campinas, Campinas, SP, Brazil}

\date{\today}

\begin{abstract}
    The Minkowski vacuum $|0\rangle_M$, which for an inertial observer is devoid of particles, is treated as a thermal bath by Rindler observers living in a single Rindler wedge \cite{unruh} as a result of the discrepancy in the definition of positive frequency between the two classes of observers and a strong entanglement between degrees of freedom in the left and right Rindler wedges. We revisit, in the context of a free scalar Klein-Gordon field, the problem of quantification of the correlations between an inertial observer Alice and left/right Rindler observes Rob/AntiRob, previously studied in \cite{martin-martinez,datta}. We emphasize the analysis of informational quantities like the locally accessible and locally inaccessible information \cite{koashi-winter,marcos-cesar,marcos-cesar2} and a closely associated entanglement measure, the entanglement of formation. We conclude that, with respect to the correlation structure probed by inertial observers alone, the introduction of a Rindler observer gives rise to a correlation redistribution which can be quantified by the entanglement of formation.
\end{abstract}

%\keywords{}

\maketitle

\section{Introduction}

The Unruh effect is one of the most important results in Quantum Field Theory in Curved Spacetimes, showing very clearly that the idea of particle is not really  a fundamental concept, being observer-dependent \cite{unruh} -- the Minkowski vacuum $|0\rangle_M$ which, by definition, is devoid of particles for an inertial observer, is experienced as a thermal bath by one uniformly accelerated observer. The reason for that is twofold: first, such uniformly accelerated has a definition of energy, and hence of positive-frequency, inequivalent to that of the inertial observer. Second, such observer experiences a causal horizon which implies an information loss that renders the pure state $|0\rangle_M$ a mixed thermal state.

\begin{figure}[ht]
\includegraphics[width=8cm]{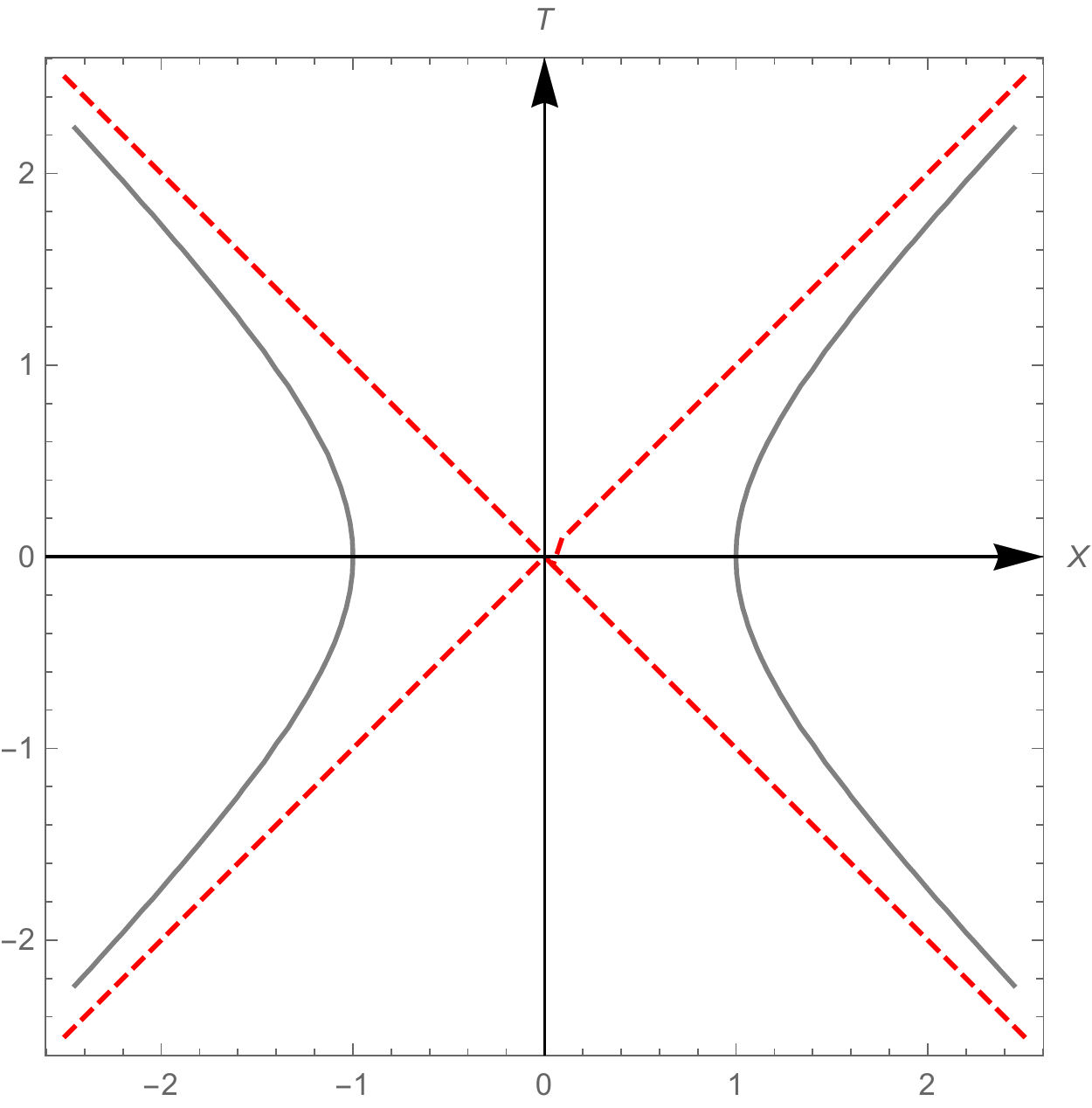}
\centering
\caption{Worldlines of Rindler observers on the right and left wedges shown in grey   (solid) lines. The Rindler horizons separating the wedges are shown as the red (dashed)  lines.}
\label{fig:rindler-space-observers}
\end{figure}

The setting to study the Unruh effect is to consider a $D$-dimensional Minkowski spacetime with coordinates $(t,x,y_i)$ and to consider observers which uniformly accelerate either towards the positive $x$ direction or towards the negative $x$ direction. The ones accelerating towards $x > 0$ are restricted to the so-called right Rindler wedge, $\mathcal{U}_{\rm I}$, defined by $x > |t|$ whereas the ones accelerating towards $x < 0$ are restricted to the so-called left Rindler wedge, $\mathcal{U}_{\rm II}$, defined by $x < -|t|$. Both classes of observers are called Rindler observers  and some of their worldlines are shown in Fig. (\ref{fig:rindler-space-observers}), where it is clear that the right and left Rindler wedges are causally disconnected from each other \footnote{In fact, the surfaces characterized by $t = \pm x $ are null surfaces respectively called past and future Rindler horizons, denoted $\mathcal{H}^-$ and $\mathcal{H}^+$, and they act as spacetime boundaries for the Rindler observers.}.

The origin of the Unruh effect lies in a strong entanglement between degrees of freedom in $\mathcal{U}_{\rm I}$ and $\mathcal{U}_{\rm II}$. In fact, it can be shown \cite{navarro-salas} that the Minkowski vacuum can be written in a basis appropriate to the Rindler observers as
\begin{equation}\label{eq:minkowski-vacuum-rindler}
    |0\rangle_M = \prod_{\omega}\dfrac{1}{\cosh \alpha_\omega}\sum_{n_\omega=0}^{\infty}\tanh^{n_\omega}\alpha_\omega |n_\omega\rangle^I |n_\omega\rangle^{II},
\end{equation}
where one has introduced the so-called \textit{squeezing parameter} $\alpha_\omega$ for each frequency by the relation
\begin{equation}\label{eq:squeezing-parameter-definition}
    \tanh \alpha_\omega = e^{-\pi\omega/a},
\end{equation}
being $a$ the acceleration parameter of the Rindler observers.
In that case, the Rindler observer is restricted to the right Rindler wedge, and thus is causally forbidden to have knowledge about the degrees of freedom in the left Rindler wedge. Due to the strong entanglement between both regions, this lack of knowledge manifests as a high degree of mixedness for the state locally probed.
The different settings where this phenomenon manifest makes it interesting for investigation from a quantum information perspective. Indeed, there are already some relevant discussions on this subject in the literature, such as Refs. \cite{martin-martinez} and \cite{datta}, where the Logarithmic Negativity was employed to qualitatively describe the entanglement in the system \cite{martin-martinez} and the Quantum Discord was numerically calculated to quantify the presence of an arbitrary quantum correlation \cite{datta}.
 Nonetheless, it would be particularly interesting that the  entanglement monotones employed to have an informational meaning behind its definition and to allow extension to further understand open problems related to the description of the physics and description of strongly accelerated bodies.
 
In this paper we compute both the classical and quantum correlation distribution as locally probed by Alice, who is in an inertial reference frame, and by Rob in the right Rindler wedge, or when probed by  Alice  and AntiRob, in the left Rindler wedge, and interpret these results in terms of locally accessible and locally inaccessible information \cite{marcos-cesar,marcos-cesar2}. This contrasts with \cite{datta} which analyzed just quantum discord for the Alice and Rob bipartition, aiming at answering the question whether or not there are quantum correlations in the near-horizon limit.
We further combine the results to the methods of \cite{koashi-winter,marcos-cesar,marcos-cesar2} which enables the computation of \textit{the entanglement of formation} for the subsystem probed by Rob and AntiRob. The ideas of \cite{koashi-winter,marcos-cesar,marcos-cesar2} allow for an interpretation of this entanglement of formation in terms of correlation redistribution. 
For each measure we have analyzed (mutual information, classical correlations, quantum discord and entanglement of formation) we review its definition and significance and give the plots against acceleration together with the way we interpret it.
The paper is organized as follows. In Sec. II we introduce the problem appropriately and review the previous treatments. In Sec. III we develop the basic canonical quantization rules in a curved spacetime. In Sec. IV we review the relevant modes for the analysis we wish to develop, mainly the Minkowski, Unruh and Rindler Modes. In Sec. V we review the transformation of the states we wish to study to the Rindler basis. In Sec. VI we present and discuss the mutual information, a result already known from \cite{martin-martinez}. In Sec. VII we discuss locally accessible and locally inaccessible informations and plot all such correlation measures together to show the correlation redistribution. In Sec. VIII we present the entanglement of formation and its role as a quantifier of correlation redistribution. Finally in Sec. IX we give the final remarks and conclusions.

\section{Observers and reference frames}

%In this subsection we review %the setup, which is the same %employed in %\cite{martin-martinez,datta}. 

Two inertial observers, Alice and Bob, observe two Unruh modes $u_i$ and $u_j$ of a massless Klein-Gordon field. Alice is supposed to carry a detector sensitive only to the frequency $\omega_i$ of mode $u_i$ whereas Bob is supposed to carry a detector sensitive only to frequency $\omega_j$ of mode $u_j$. The state of the system is supposed to be a maximally entangled state
\begin{equation}
    \label{eq:state-inertial-observers}
    |\psi\rangle = \frac{1}{\sqrt{2}}\bigg(|0_i\rangle_M|0_j\rangle_M + |1_i\rangle_M |1_j\rangle_M\bigg).
\end{equation}
It is fundamental to understand here that this is a bipartition with respect to modes. The fact that Alice's detector can only detect frequency $\omega_i$ and not $\omega_j$ implies that she has no access to the subsystem that Bob has access and \textit{vice-versa.} In particular, any information Alice is able to gather about the mode $u_j$ is through correlations.

One now introduces a Rindler observer, living on the right Rindler wedge $\mathcal{U}_I$, Rob, which also carries a detector sensitive only to frequency $\omega_j$. It is fundamental here that one is working with Unruh modes. In that case, a mode which for the inertial observer has some frequency $\omega$, has the same frequency $\omega$ for the Rindler observer \cite{martin-martinez}. 
In that setting, inasmuch as Bob, Rob only observes mode $u_j$. But since we have the two disconnected Rindler wedges, $\mathcal{U}_I$ and $\mathcal{U}_{II}$, isolated from each other by causal horizons, the transformation from the inertial basis to the Rindler basis introduces one additional bipartition between the two regions, c.f. Eq. (\ref{eq:minkowski-vacuum-rindler}).
While Rob can only access the part of the state of mode $u_j$ associated to $\mathcal{U}_I$ a symmetric Rindler observer on the left Rindler wedge $\mathcal{U}_{II}$ can observe the complementary part. The introduction of said observer has been done in Ref. \cite{martin-martinez}, where he is conventionally called AntiRob. Hence the setup is that of a tripartite quantum state observed by three observers: Alice, Rob and AntiRob, the first being inertial, and the other two being uniformly accelerated, living respectively on regions $I$ and $II$. Alice carries a detector sensitive only to mode $u_i$ whereas Rob and AntiRob carry detectors sensitive only to mode $u_j$.
In that sense, the state, which when probed by Alice and Bob was naturally bipartite - a bipartition between modes $u_i$ and $u_j$, now when probed by Alice, Rob and AntiRob is naturally tripartite - a bipartition between modes $u_i$ and $u_j$ with a second bipartition between regions $\mathcal{U}_I$ and $\mathcal{U}_{II}$ affecting the modes $u_j$ subsystem. That is the setup of Ref. \cite{martin-martinez,datta} which we consider.

In \cite{martin-martinez}, the author considered the evaluation of \textit{the mutual information} of the three possible bipartite subsystems and \textit{the negativity} as a measure of entanglement. The negativity of the bipartitions between Alice-Rob and Alice-AntiRob was seem to decrease to zero in the infinite acceleration limit, which corresponds to the near-horizon limit. The authors concluded that an entanglement degradation existed due to the horizon and supposed that there remain no quantum correlations on the near-horizon limit, since the negativity vanishes in said situation. Moreover the authors found that the negativity for the bipartition between Rob-AntiRob, i.e., between the observers separated by the causal horizon, increased in the near-horizon limit. They noticed, however, that this entanglement is not useful as a resource because these two observers are forbidden classical communication.
In \cite{datta}, the author evaluated \textit{the quantum discord} of the state probed by Alice and Rob. The main conclusion was a comparison to the negativity and the observation that even though the negativity vanishes on the near-horizon limit, the quantum discord does not, signaling that there are still quantum correlations in that limit, which are not captured by the negativity.

\section{Basic Canonical Quantization in Curved Spacetimes}

We shall review the most basic approach to the canonical quantization of a Klein-Gordon field in a globally hyperbolic spacetime. We just give the basic definitions and results, referring the reader to \cite{birrell-davies,wald-qft} for a complete account of the subject.

Let $(M,g)$ be a globally hyperbolic spacetime and $\phi$ be a real, minimally coupled Klein-Gordon field on said background \cite{wald}. The dynamics of $\phi$ can be encoded in the Lagrangian density
\begin{equation}
    \mathcal{L}=\frac{1}{2}\left(\nabla_\mu \phi \nabla^\mu \phi -m^2 \phi^2\right)\sqrt{|g|}.
\end{equation}
The equation of motion deriving from this Lagrangian is the well-known Klein-Gordon equation
\begin{equation}\label{eq:kg-equation}
    (\Box - m^2)\phi = 0.
\end{equation}
Given a Cauchy surface $\Sigma\subset M$, with normal vector field $n^\mu$ and induced metric $h$, we can describe this system in the canonical formalism \cite{wald-qft}. In particular, the $\phi(x)$, for $x\in \Sigma$, act as the coordinates, and the conjugate momenta are $\pi(x)$, given by \cite{wald-qft}:
\begin{equation}
    \pi(x) = (n^\mu(x)\nabla_\mu \phi(x))\sqrt{h}.
\end{equation}
Canonical quantization therefore amount to finding a Hilbert space on which operators $\phi(x),\pi(x)$ act satsifying the equal-time canonical commutation relations
\begin{equation}\label{eq:ccr}
    [\phi(x),\pi(y)]=i\delta(x,y),\quad x,y\in \Sigma.
\end{equation}
A manner of doing so is to expand the field into modes. In particular, recall that we can define, on the space of solutions to the Klein-Gordon equation, the bilinear form
\begin{equation}
    (\phi,\psi)=i\int_{\Sigma}(\phi^\ast \nabla_\mu \psi-\psi\nabla_\mu \phi^\ast)n^\mu d\Sigma,
\end{equation}
which satisfies all axioms of an inner product except that it is not positive-definite. One then introduces a set of the so-called mode functions $\{u_i,u_i^\ast\}$ such that
\begin{equation}\label{eq:orthogonality-relations}
    (u_i,u_j)=\delta_{ij},\quad (u_i,u_j^\ast)=0,\quad (u_i^\ast,u_j^\ast)=-\delta_{ij},
\end{equation}
and such that any real solution $\phi$ to Eq. (\ref{eq:kg-equation}) can be uniquely expanded as
\begin{equation}
    \phi = \sum_i a_i u_i + a_i^\ast u_i^\ast.
\end{equation}

Turning $\phi(x)$ to operators would then amount to turning $a_i$ to operators. One may further argue \cite{birrell-davies} that the canonical commutation relations (\ref{eq:ccr}) are equivalent to the commutation relations
\begin{equation}
    [a_i,a_j^\dagger]=\delta_{ij},\quad [a_i,a_j]=[a_i^\dagger,a_j^\dagger].
\end{equation}
This justifies a Fock space picture on which we have a vacuum state $|0\rangle$ defined by 
\begin{equation}a_i|0\rangle = 0.\end{equation} Being more explicit, the one-particle space is taken as the Hilbert space completion of the space of solutions spanned by just the $\{u_i\}$, without the complex conjugates, with inner product being the restriction of the Klein-Gordon form $(,)$ to that subspace. The Hilbert space of the theory is the bosonic Fock space construct based on said one-particle Hilbert space \cite{wald-qft,wald}. 
The interpretation of the construction lies in the observation that if the $u_i$ are positive-frequency with respect to a family of observers following the integral lines of a timelike future-directed normalized vector field $Z$, meaning that there are $\omega_i\in [0,+\infty)$ satisfying
\begin{equation}
    \mathfrak{L}_Z u_i = -i\omega_i u_i,\quad \omega_i > 0,
\end{equation}
then the states $|1_i\rangle = a_i^\dagger |0\rangle$ are states containing just one particle with energy $\omega_i$.

In general there is not a preferred choice of mode functions $\{u_i,u_i^\ast\}$ as there is no privileged notion of time in an arbitrary spacetime. It is important therefore to relate the constructions obtained by two distinct choices of mode functions $\{u_i,u_i^\ast\}$ and $\{\bar{u}_i,\bar{u}_i^\ast\}$. This may be accomplished by expanding one set in terms of the other set:
\begin{equation}\label{eq:bogolubov-expansion}
    \bar{u}_i = \sum_j \alpha_{ij}u_j + \beta_{ij}u_j^\ast,
\end{equation}
in terms of the Bogolyubov coefficients $\alpha_{ij},\beta_{ij}$ \cite{birrell-davies}. These coefficients, which can be straightforwardly computed using Eq. (\ref{eq:orthogonality-relations}) to be
\begin{equation}
    \alpha_{ij}=(u_j,\bar{u}_i),\quad \beta_{ij}=-(u_j^\ast,\bar{u}_i),
\end{equation}
allow to establish a relation between the annihilation and creation operators $a_i,a_i^\dagger$ of the first quantization and $\bar{a}_i,\bar{a}_i^\dagger$ of the second quantization. The relation, which, in this approach, is the central aspect of the comparisson of the two quantizations, is given by
\begin{equation}\label{eq:transformation-ladder-operators}
    \bar{a}_i = \sum_j \alpha_{ij}a_i-\beta_{ij}a_i^\dagger,\quad \bar{a}_i^\dagger = \sum_j \alpha_{ij}^\ast a_i^\dagger - \beta_{ij}^\ast a_i.
\end{equation}
It immediately follows from Eq. (\ref{eq:transformation-ladder-operators}) that if $|0\rangle$ is the vacuum of the first quantization then, in general, $\bar{a}_i|0\rangle\neq 0$, so that it \textit{does not coincide} with the vacuum $|\bar{0}\rangle$ of the second quantization. The two vacua coincide if and only if $\beta_{ij}=0$, which in turn happens when each $\bar{u}_i$ is a combination of only positive frequency modes of the first quantization.
This is the straightforward treatment taken in many references, for instance in \cite{birrell-davies}. A more rigorous treatment is presented in \cite{wald-qft}.

\section{Minkowski, Unruh and Rindler Modes}

We now briefly review the modes of importance for the analysis we wish to develop. Consider now the two-dimensional Minkowski spacetime $(M,\eta)$ with the flat metric $(\eta_{\mu\nu})=\operatorname{diag}(-1,1)$. Let a real massless Klein-Gordon field $\phi$ be given.
We wish to quantize the field according to the ideas of the previous section. Minkowski spacetime, however, has a timelike Killing vector field, and hence has a class of distinguished observers whose four-velocity is proportional to the Killing field, namely the inertial observers. To quantize the field in appropriate manner to this class of observers one chooses modes which are positive frequency with respect to the inertial observers. A natural family of such modes are the plane waves
\begin{equation}
    u^M_k(t,x) = \frac{1}{\sqrt{4\pi\omega_k}}e^{-i\omega t + ikx},\quad \omega_k=|k|, k\in \mathbb{R},
\end{equation}
where the superscript $M$ indicates these are Minkowski modes.
These modes define the usual Minkowski quantization used in most flat-spacetime Quantum Field Theory textbooks \cite{duncan,weinberg,peskin,schwartz}. They are labelled by a real number $k\in \mathbb{R}$ and divide in two classes: the left-moving solutions with $k < 0$ and the right-moving solutions with $k > 0$. Employing these modes, the field expands as
\begin{equation}
    \phi(t,x)=\int_{\mathbb{R}} \bigg(u_k^M(t,x)a(k)+{u_k^M}^\ast(t,x)a^\dagger(k) \bigg)dk,
\end{equation}
and it is evident that decomposing the integral on the regions $k < 0$ and $k > 0$ we may write
\begin{equation}
    \phi(t,x) = \phi_-(t,x)+\phi_+(t,x),
\end{equation}
where $\phi_-$ and $\phi_+$ are, respectively, the parts of the integral with $k < 0$ and $k > 0$. This equation tells that in the quantum theory the two sectors decouple and can be studied independently \cite{carroll}.

Now let us consider one eternally uniformly accelerating observer, also known as a Rindler observer, living in the right Rindler wedge $\mathcal{U}_I$ defined in the introduction.
We can choose coordinates $(\eta,\xi)$ on $\mathcal{U}_I$ adapted to the worldlines of such observers, given by
\begin{equation}\label{eq:minkowski-rindler-1-transformation}
    t = \frac{1}{a}e^{a\xi}\sinh(a\eta), \quad z = \frac{1}{a}e^{a\xi}\cosh(a\eta).
\end{equation}
In terms of these coordinates the appropriate modes for the right Rindler observer are the modes
\begin{equation}\label{eq:right-rindler-modes}
    u^{(I)}_k(\eta,\xi)=\dfrac{1}{\sqrt{4\pi \omega_k}}e^{-i\omega \eta+ik\xi},\quad \omega_k = |k|,k \in \mathbb{R}.
\end{equation}
where now the superscript $I$ indicates these are Rindler modes on the right Rindler wedge.

We wish to compare the two quantizations, but the Minkowski observer also probes degrees of freedom in the left Rindler wedge. Because of that, we also need to consider Rindler modes for a Rindler observer supported in that region. We similarly introduce in region $\mathcal{U}_{II}$ coordinates $(\eta,\xi)$, denoted by the same symbols as those in region $\mathcal{U}_I$, which relate to the Minkowski coordinates by
\begin{equation}\label{eq:minkowski-rindler-2-transformation}
     t = -\frac{1}{a}e^{a\xi}\sinh(a\eta), \quad z = -\frac{1}{a}e^{a\xi}\cosh(a\eta).
\end{equation}
Using these coordinates the appropriate mode functions for the left Rindler observer are the modes
\begin{equation}\label{eq:left-rindler-modes}
    u_k^{II}(\eta,\xi)=\dfrac{1}{\sqrt{4\pi \omega_k}}e^{i\omega_k \eta+ik\xi},\quad \omega_k=|k|,k\in \mathbb{R}.
\end{equation}

The modes $u_k^I$ and $u_k^{II}$ are at first defined just on $\mathcal{U}_I$ and $\mathcal{U}_{II}$, but if we set $u_k^I$ to be zero on $\mathcal{U}_{II}$ and $u_k^{II}$ to be zero on $\mathcal{U}_{I}$, then the set of modes $\{u_k^{(I)},{u_k^{(I)}}^\ast,u_k^{(II)},{u_k^{(II)}}^\ast\}$ taken together allow us to expand the general solution to the Klein-Gordon equation as
    \begin{equation}
        \phi(\eta,\xi) = \sum_{\Omega=I,II}\int_\mathbb{R} \bigg(b^\Omega(k)u_k^\Omega(\eta,\xi)+{b^\Omega}^\dagger(k){u_k^\Omega}^\ast(\eta,\xi)\bigg)dk.
    \end{equation}
If we consider the space of solutions spanned by just the positive-frequency modes $\{u_k^I,u_k^{II}\}$ then it is clear that its Hilbert space completion is a direct sum $\mathfrak{H}_I\oplus \mathfrak{H}_{II}$ where $\mathfrak{H}_I$ is the space spanned by just the $\{u_k^I\}$ and $\mathfrak{H}_{II}$ is the space spanned by just the $\{u_k^{II}\}$. It follows immediately that, for the Rindler quantization, the Fock space is a tensor product \begin{equation}\mathcal{F}(\mathfrak{H}_I\oplus \mathfrak{H}_{II})=\mathcal{F}(\mathfrak{H}_I)\otimes \mathcal{F}(\mathfrak{H}_{II}).\end{equation} In particular the Rindler vacuum $|0\rangle^R$ decomposes as a product
\begin{equation}
    |0\rangle^R = |0\rangle^I\otimes |0\rangle^{II}.
\end{equation}

We wish to compare the quantizations, which would entail the computation of the Bogolyubov coefficients relating the Rindler modes and Minkowski modes. There is, however, a workaround, which will be further relevant for the informational analysis. The idea, due to Unruh \cite{unruh}, is to find a new set of modes which reproduces the Minkowski quantization, but which transform more simply in terms of the Rindler modes \cite{martin-martinez}.
The appropriate modes are the so-called Unruh modes, defined by
\begin{subequations}
\begin{align}\label{eq:unruh-modes-definition}
    h_k^{I}&=\dfrac{1}{\sqrt{2\sinh\left(\frac{\pi\omega}{a}\right)}}\left(e^{\pi\omega/2a}u_k^{I}+e^{-\pi\omega/2a}u_{-k}^{II}\right),\\
     h_{k}^{II}&=\dfrac{1}{\sqrt{2\sinh\left(\frac{\pi\omega}{a}\right)}}\left(e^{\pi\omega/2a}u_k^{II}+e^{-\pi\omega/2a}u_{-k}^{I}\right).
\end{align}
\end{subequations}
These modes allows for an expansion of the general solution to the Klein-Gordon equation, which in the quantum theory gives rise to the annihilation and creation operators
\begin{equation}
    \phi(t,x)=\sum_{\Omega=I,II}\int \bigg(c_k^\Omega h_k^\Omega(t,x)+{c_k^\Omega}^\dagger {h_k^\Omega}^\ast(t,x)\bigg) dk.
\end{equation}

It is important to remark here that the quantization  based on the Unruh modes is equivalent to the Minkowski quantization. Moreover, the Unruh modes are labelled by superscripts $I$ and $II$. They are not related to the Rindler wedges, but to the left-moving and right-moving sectors of the Minkowski quantization. When Unruh modes are employed, the right-moving sector corresponds to the sector spanned by just the $\{h_k^I,{h_k^{I}}^\ast\}$, whereas the left-moving sector corresponds to the sector spanned by just the $\{h_k^{II},{h_k^{II}}^\ast\}$. Since, as already mentioned, the left-moving and right-moving sectors decouple, we henceforth focus just on the right-moving sector, which entails in Minkowski quantization to consider just $k > 0$ and in Unruh modes quantization to consider just the $\{h_k^{I},{h_k^{I}}^\ast\}$.

The Unruh modes are defined explicitly in terms of Rindler modes in Eqs. (\ref{eq:unruh-modes-definition}) and so the Bogolyubov coefficients relating $h_k^I,{h_k^{I}}^\ast$ with the Rindler modes can be obtained by inspection \cite{carroll}, being simply:
\begin{subequations}\label{eq:unruh-modes-bogolubov}
\begin{align}\alpha_{kk'}^{(I)}&=\cosh \alpha_\omega \delta(k-k'),\;\; \alpha_{kk'}^{(II)}=0,\\ \beta_{kk'}^{(I)}&=0,\;\; \beta_{kk'}^{(II)}=\sinh\alpha_\omega \delta(k-k'),
\end{align}
\end{subequations}
where one has introduced the squeezing parameter $\alpha_\omega$ and where $\omega = |k|$ as already explained.
Since the Unruh quantization agrees with the Minkowski one, they share the same vacuum $|0\rangle^M$ and the Minkowski vacuum may be defined by the equation \begin{equation}\label{eq:minkowski-vacuum-equation}
    c_k^I |0\rangle^M = c_k^{II}|0\rangle^M = 0,
\end{equation} for $k\in\mathbb{R}$. Using Eqs. (\ref{eq:unruh-modes-bogolubov}) to express $c_k^I,c_k^{II}$ in terms of $b_k^I,{b_k^I}^\dagger,b_k^{II},{b_k^{II}}^\dagger$, one is able to solve Eq. (\ref{eq:minkowski-vacuum-equation}) on the Rindler basis and express the Minkowski vacuum in a basis meaningful for the Rindler observers. When this is done one obtains Eq. (\ref{eq:minkowski-vacuum-rindler}) \cite{navarro-salas}.
%, which we reproduce here for benefit of the reader
%\begin{equation}
%    |0\rangle_M = \prod_{\omega}\dfrac{1}{\cosh %\alpha_\omega}\sum_{n_\omega=0}^{\infty}\tanh^n\alpha_\omega %|n_\omega\rangle^I |n_\omega\rangle^{II}.
%\end{equation}
To find out what a Rindler observer living in the right Rindler wedge $\mathcal{U}_I$ perceives one would trace out the modes supported in region $\mathcal{U}_{II}$. If this is done one obtains the density operator
\begin{equation}
    \rho_I = \prod_\omega \frac{1}{\cosh^2\alpha_\omega}\sum_{n_\omega=0}^\infty \tanh^{2n}\alpha_\omega |n_\omega\rangle^I\langle n_\omega|.
\end{equation}
Upon recalling the definition of the squeezing parameter, Eq. (\ref{eq:squeezing-parameter-definition}), one is able to find out that this is, in fact, a \textit{thermal density operator} with temperature, in natural units,
$T = \frac{a}{2\pi}$.
This result is the Unruh effect, stating that the Minkowski vacuum is perceived as a thermal mixed state for Rindler observers living in either the right or left Rindler wedges.

%\section{Relativistic Quantum Information}

\section{States in the Rindler Basis}
Following previous works on the subject \cite{martin-martinez,datta}, the first step to perform the analysis is to expand the mode $u_j$ part of the state in Eq. (\ref{eq:state-inertial-observers}) into a basis appropriate for the Rindler observers. This can be done employing the transformation of the Minkowski vacuum, Eq. (\ref{eq:minkowski-vacuum-rindler}), together with the transformation of the creation and annihilation operators of Unruh modes to creation and annihilation operators of Rindler modes, which can be done with the general transformation, Eq. (\ref{eq:transformation-ladder-operators}), together with the concrete Bogolyubov coefficients, Eq. (\ref{eq:unruh-modes-bogolubov}). This last step gives the one-particle Unruh state:
\begin{equation}\label{eq:4-unruh-one-particle-rindler}
    |1\rangle_j^U = \dfrac{1}{\cosh^2 \alpha_j}\sum_{n=0}^\infty \tanh^{n}\alpha_j \sqrt{n+1}|n+1\rangle_j^{I} |n\rangle^{II}_j,
\end{equation}
where $\alpha_j$ is the squeezing parameter associated to the frequency $\omega_j$ of mode $u_j$. The fact that the transformation does not change the frequency, but only the occupation numbers, leading a monochromatic state to another monochromatic state, is the reason to use Unruh modes for this analysis.
With this data, following the notation of \cite{martin-martinez}, let $\rho_{AR},\rho_{A\bar{R}},\rho_{R\bar{R}}$ be the bipartite states of the subsystems probed by Alice-Rob, Alice-AntiRob and Rob-AntiRob. Employing the above transformations, the states in the Rindler basis are seem to be
\begin{widetext}
    \begin{eqnarray}
            \rho_{AR} &=& \dfrac{1}{2\cosh^2\alpha_j}\sum_{n=0}^\infty \tanh^{2n}\alpha_j \left[|0n\rangle_{ij}^{MI}\langle 0n|+\dfrac{(n+1)}{\cosh^2\alpha_j}|1n+1\rangle_{ij}^{MI}\langle 1n+1|\right.\nonumber\\ &&\left.+\dfrac{\sqrt{n+1}}{\cosh\alpha_j} \bigg(|0n\rangle_{ij}^{MI}\langle 1n+1|+|1n+1\rangle_{ij}^{MI}\langle 0n|\bigg)\right],\label{eq:state-alice-rob}\\
            \rho_{A\bar{R}}&=&\dfrac{1}{2\cosh^2\alpha_j}\sum_{n=0}^\infty \tanh^{2n}\alpha_j \left[|0n\rangle_{ij}^{M,II}\langle 0n|+\dfrac{n+1}{\cosh^2\alpha_j}|1n\rangle_{ij}^{M,II}\langle 1n|\right. \nonumber\\ &&\left.+\dfrac{\sqrt{n+1}\tanh\alpha_j}{\cosh\alpha_j} \bigg(|0n+1\rangle_{ij}^{M,II}\langle 1n|+|1n\rangle_{ij}^{M,II}\langle 0n+1|\bigg)\right],\label{eq:state-alice-antirob}\\ 
            \rho_{R\bar{R}}&=&\dfrac{1}{2\cosh^2\alpha_j}\sum_{n,m=0}^\infty \tanh^{n+m}\alpha_j \bigg(|nn\rangle_{j}^{I,II}\langle mm|+\dfrac{\sqrt{n+1}\sqrt{m+1}}{\cosh^2\alpha_j}|n+1n\rangle_j^{I,II}\langle m+1 m|\bigg)\label{eq:state-rob-antirob}.
    \end{eqnarray}
\end{widetext}

In that same way, one may further obtain the states of the three subsystems probed individually by Alice, Rob and AntiRob. It gives
    \begin{equation}
        \rho_A = \frac{1}{2}(|0\rangle_i^M\langle 0|+|1\rangle_i^M\langle 1|)\label{eq:state-alice},\end{equation}\begin{equation}
        \rho_R = \sum_{n=0}^\infty \dfrac{\tanh^{2(n-1)}\alpha_j}{2\cosh^2\alpha_j}\bigg(\tanh^2\alpha_j +\frac{n}{\cosh^2\alpha_j}\bigg)|n\rangle_j^I\langle n|\label{eq:state-rob},\end{equation} \begin{equation}
        \rho_{\bar{R}}= \dfrac{1}{2\cosh^2\alpha_j}\sum_{n=0}^\infty \tanh^{2n}\alpha_j \bigg(1+\frac{n+1}{\cosh^2\alpha_j}\bigg)|n\rangle_j^{II}\langle n|\label{eq:state-antirob}.\end{equation}

\section{Mutual Information}
Let $\rho_{AB}$ be a bipartite state and let $\rho_A = \operatorname{Tr}_B\rho_{AB}$ and $\rho_B = \operatorname{Tr}_A \rho_{AB}$ be the marginal states of the two subsystems, $A$ and $B$, respectively. The mutual information is defined to be \begin{equation}\label{eq:mutual-information}
    I(\rho_{AB})=S(\rho_{A})+S(\rho_{B})-S(\rho_{AB})
\end{equation}
where $S(\rho)$ is the Von-Neumann entropy of the density operator $\rho$. The mutual information quantifies the total correlations between the two parts.
If we have a tripartite pure state $\rho_{ABC} = |\psi\rangle\langle \psi|$ then we can extract three marginal bipartite states from it, 
    $\rho_{AB}=\operatorname{Tr}_C \rho_{ABC}$, 
    $\rho_{AC}=\operatorname{Tr}_B \rho_{ABC}$, and
    $\rho_{BC}=\operatorname{Tr}_A \rho_{ABC}$.
In that case, the fact that $\rho_{ABC}$ is pure implies the equalities
\begin{subequations}
    \begin{align}\label{eq:entropies-tripartite}
    S(\rho_{AB})&=S(\rho_{C}),\\
    S(\rho_{AC})&=S(\rho_{B}),\\
    S(\rho_{BC})&=S(\rho_{A}),
    \end{align}
\end{subequations}
and this implies that the evaluation of the mutual information for such bipartite states reduces to the problem of evaluating the Von-Neumman entropies $S(\rho_A),S(\rho_B),S(\rho_C)$.
\begin{figure}[ht]
\includegraphics[width=\linewidth]{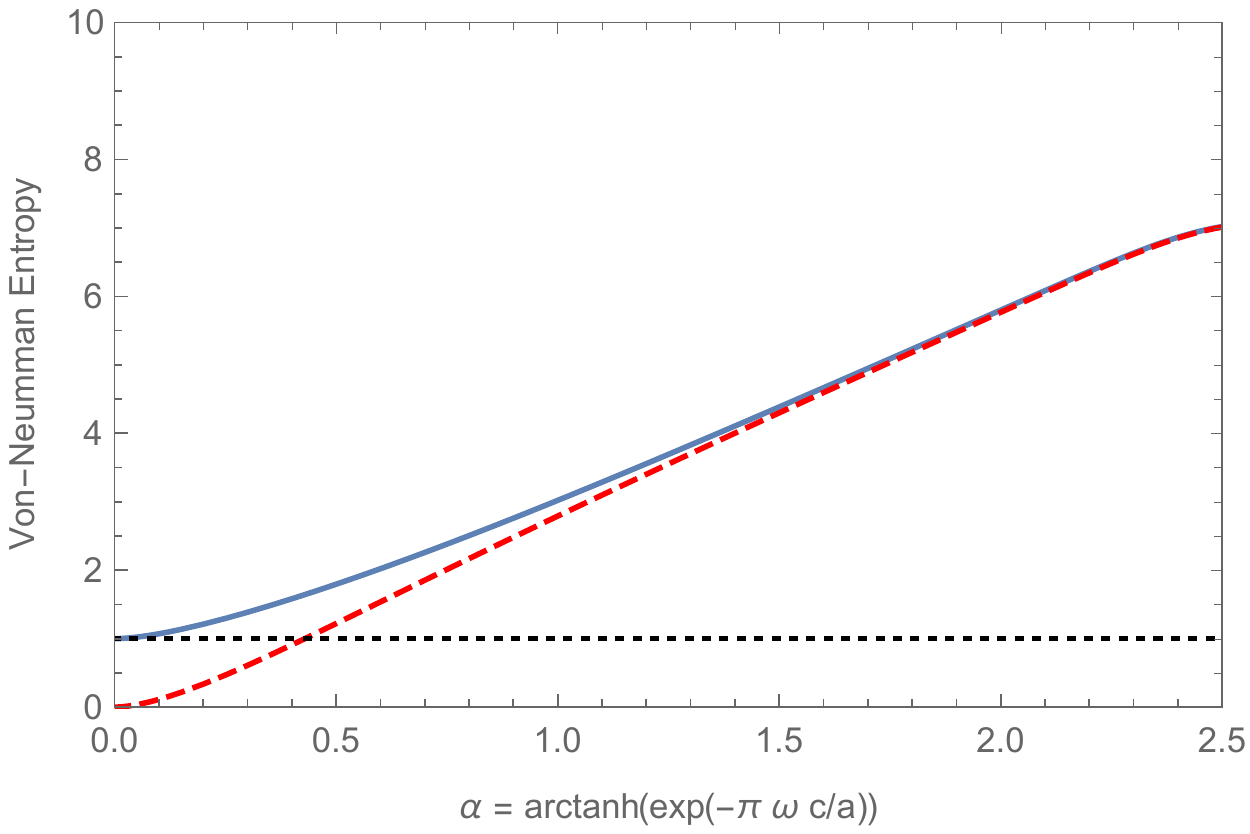}
\centering
\caption[Von-Neumman entropies of the states $\rho_A,\rho_R,\rho_{\bar{R}}$]{Von-Neumman entropies of the states $\rho_A,\rho_R,\rho_{\bar{R}}$. The state $\rho_A$ is shown by the black (dotted) line, $\rho_R$ by the blue (solid) line and $\rho_{\bar{R}}$ by the red (dashed) line.}
\label{fig:vonNeumman-entropies-three-states-squeeze}
\end{figure}

\begin{figure}[ht]
\includegraphics[width=\linewidth]{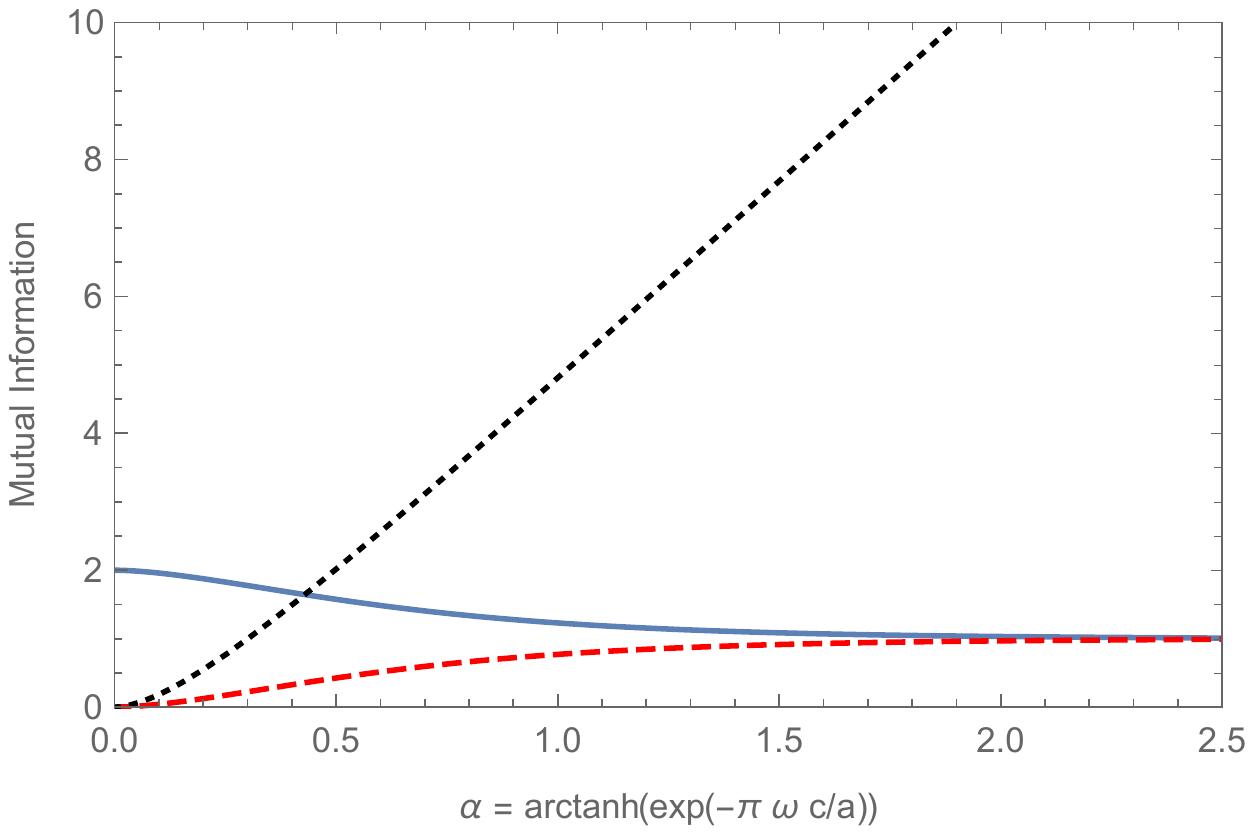}
\centering
\caption[Mutual information of the states $\rho_{AR},\rho_{A\bar{R}}$ and $\rho_{R\bar{R}}$]{Mutual information - The state $\rho_{AR}$ is the blue solid line, $\rho_{A\bar{R}}$ is the red dashed line and $\rho_{R\bar{R}}$ is the black dotted line.}
\label{fig:mutual-information-three-states-squeeze}
\end{figure}

For the concrete problem we considered, all the states $\rho_A,\rho_R,\rho_{\bar{R}}$ of the individual subsystems are diagonal in the Rindler occupation number basis. This means that evaluating the Von-Neumman entropies can be done straightforwardly. The results are shown in Fig. (\ref{fig:vonNeumman-entropies-three-states-squeeze}).
From these entropies the mutual informations can be straightforwardly obtained as well. The result is shown in Fig. (\ref{fig:mutual-information-three-states-squeeze}).
In Fig. (\ref{fig:mutual-information-three-states-squeeze}) we see that the higher the squeezing parameter the more the correlations of the bipartition between the Minkowski and Rindler observer on the right Rindler wedge decrease and the more the correlations between the Minkowski and Rindler observer on the left Rindler wedge increase. Moreover, this follows a conservation law \cite{martin-martinez}. In fact, we have from Eqs. (\ref{eq:mutual-information}) and (\ref{eq:entropies-tripartite}),
\begin{equation}
    I(\rho_{AR})+I(\rho_{A\bar{R}})=2S(\rho_A),
\end{equation}
but it follows immediately from Eq. (\ref{eq:state-alice}) that $S(\rho_A)=1$ and we find
\begin{equation}
    I(\rho_{AR})+I(\rho_{A\bar{R}})=2,
\end{equation}
which works as a conservation law, which is obeyed as a correlation transfer from the bipartition $\rho_{AR}$ to the bipartition $\rho_{A\bar{R}}$. The meaning of this redistribution of correlation will become clear as we discuss the distinction of classical and quantum correlation in what follows.

\section{Locally Accessible and Inaccessible Information}
Let again $\rho_{AB}$ be a bipartite state. The mutual information quantifies the total amount of information contained in the correlations between the two parts. One may ask how much of such information is locally accessible to each part by local measurements. This is quantified by the Locally Accessible Information (LAI), also known as Classical Correlations. To define it, suppose we wish to find how much information is accessible locally to $B$ by local measurements. In that case the LAI is defined to be
\begin{equation}\label{eq:def-lai}
    J^\leftarrow(\rho_{AB})=\max_{\mathbf{1}\otimes \Pi} \bigg[S(\rho_A)-\sum_\lambda p_\lambda S(\rho_A^\lambda)\bigg],
\end{equation}
where the maximum is taken over all possible projective measurements on the $B$ subsystem. For each measurement $\Pi$ the probabilities are $p_\lambda$ and the post-selected states of the $A$ subsystem are $\rho_A^\lambda$ (in other words, one takes the post-selected state of the composite system and traces $B$ out).
In the notation the arrow points away from the system being measured. This quantity measures the maximum decrease in the uncertainty of the state of $A$ that a measurement in $B$ might impart, that being the reason why it is called Locally Accessible Information. The reason for the name Classical Correlations lies in the fact that $J^\leftarrow(\rho_{AB})$ satisfies properties that a quantifier of \textit{exclusively classical correlations} should satisfy.

Associated to the LAI there is the Locally Inaccessible Information (LII). The idea is that it should quantify the amount of information contained in the correlations which cannot be accessed locally through measurements. Since $I(\rho_{AB})$ quantifies the total correlations, this can be defined straightforwardly as
\begin{equation}\label{eq:def-lii}
    \mathcal{D}^\leftarrow(\rho_{AB})=I(\rho_{AB})-J^\leftarrow(\rho_{AB}).
\end{equation}
This quantity is also known as the quantum discord and, inasmuch as $J^\leftarrow(\rho_{AB})$ quantifies the classical part of the correlations, it quantifies the quantum part of the correlations.
It can be seem from the definitions Eqs. (\ref{eq:def-lai}) and (\ref{eq:def-lii}) that both quantities can be very hard to be computed due to the optimization they require. For a special case there is a simplification, that being the central idea of the method employed in \cite{datta}.
The idea is that if a bipartite state $\rho_{AB}$ has one part which is effectively two-level, by which we mean it is written as
\begin{equation}\label{eq:effectively-two-level}
    \rho_{AB}=\sum_{a,b=0,1}M_{ab}\otimes |a\rangle \langle b|,
\end{equation}
then one may focus on the measurements in $B$ which lie in the subspace spanned by the basis operators $|a\rangle\langle b|$ for $a,b=0,1$. In that case, the measurements are represented by $2\times 2$ projectors and this enables to parameterize them by points on a sphere. Concretely, every measurement of interest consists of two projectors $\Pi_\pm(\mathbf{x})$ for $\mathbf{x}\in S^2$ given by
\begin{equation}
    \Pi_{\pm}(\mathbf{x}) = \frac{1}{2}\left[\mathbf{1}\pm\mathbf{x}\cdot \sigma\right],\quad \mathbf{x}\in S^2,
\end{equation}
where $\sigma=(\sigma_1,\sigma_2,\sigma_3)$ is a vector whose components are three operators whose matrix representations are the Pauli matrices. In that case, $J^\leftarrow(\rho_{AB})$ may be obtained by an optimization over $S^2$ and $\mathcal{D}^\leftarrow(\rho_{AB})$ may be obtained from it.

\begin{figure}[ht]
\includegraphics[width=\linewidth]{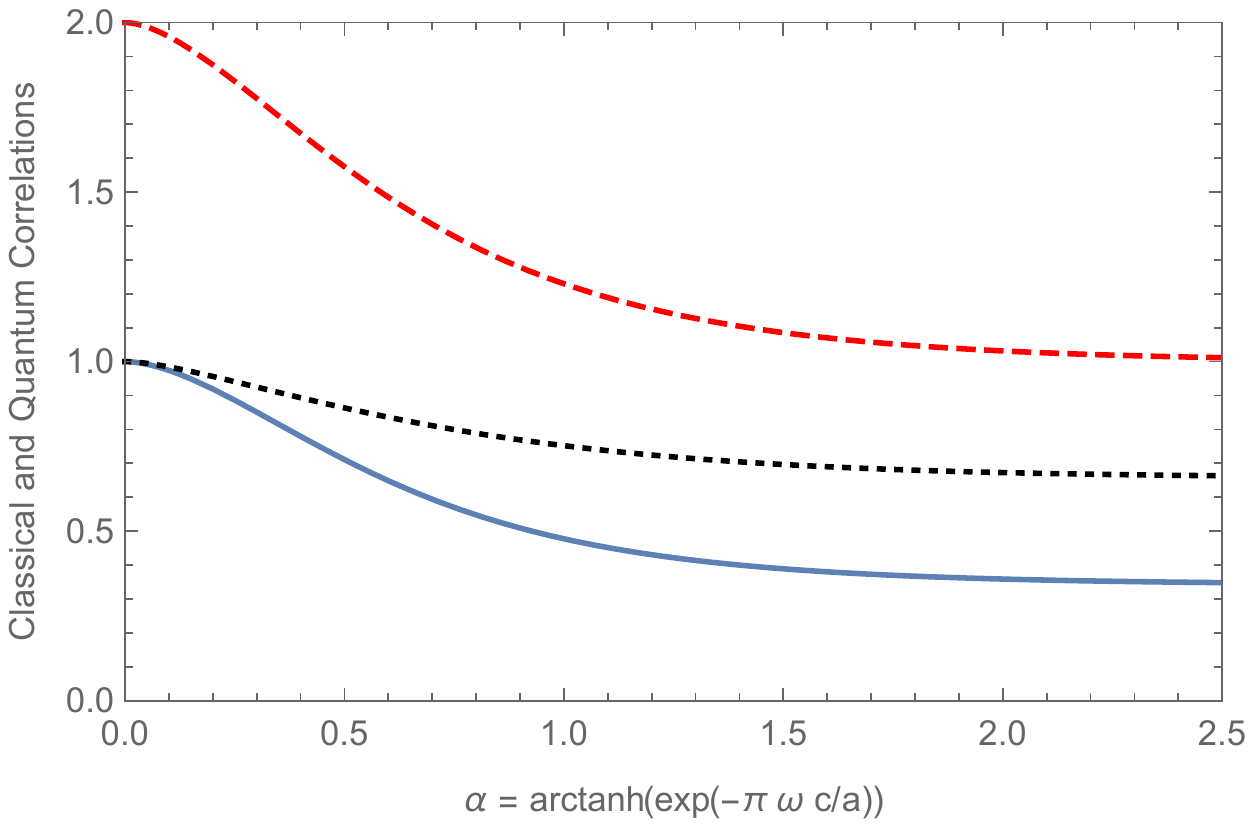}
\centering
\caption{State $\rho_{AR}$ - Classical Correlations is the solid blue line, quantum discord is the black (dotted) line and mutual information is the red (dashed) line.}
\label{fig:all-correlations-rindler-M-I-squeeze}
\end{figure}

\begin{figure}[ht]
\includegraphics[width=\linewidth]{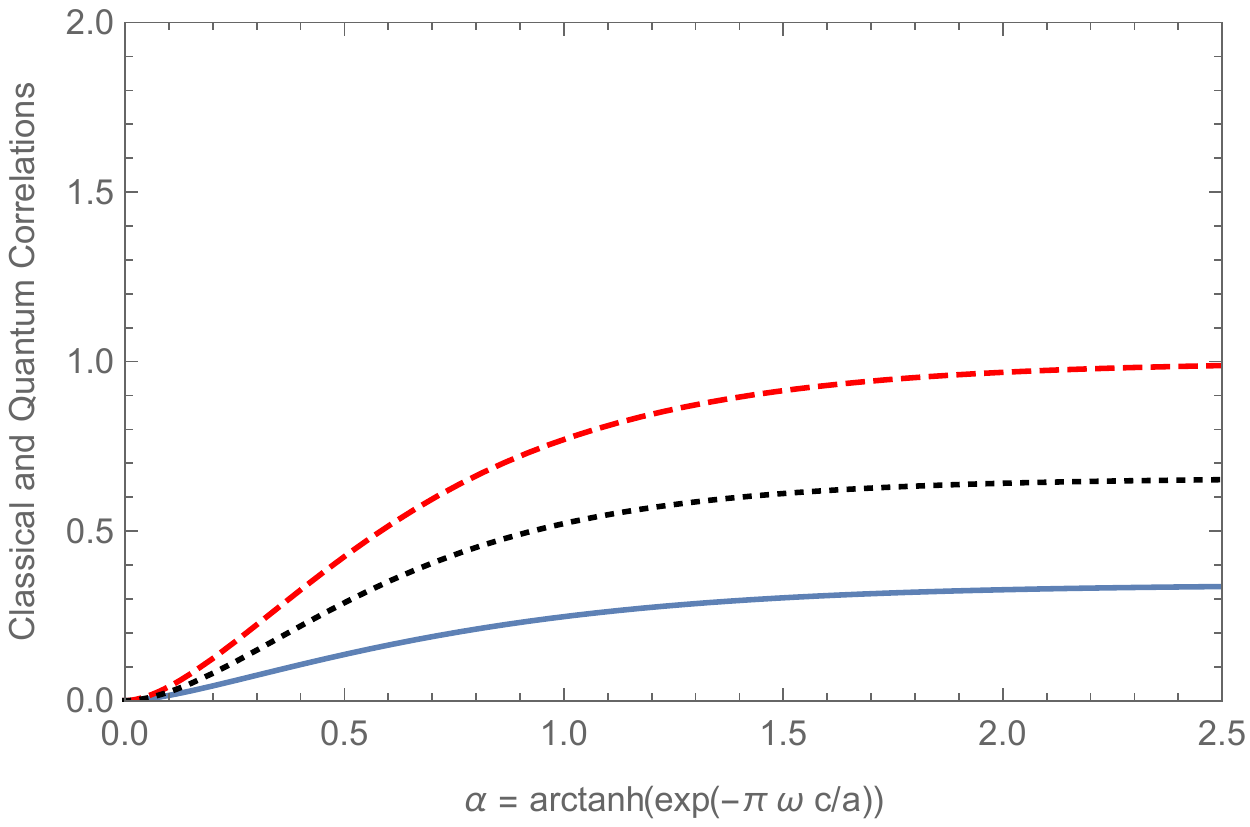}
\centering
\caption{State $\rho_{A\bar{R}}$ - Classical Correlations is the solid blue line, quantum discord is the black dotted line and mutual information is the red dashed line.}
\label{fig:all-correlations-rindler-M-II-squeeze}
\end{figure}

Considering the concrete state we are working with, there are two bipartitions to which the method applies, namely the bipartitions between Alice-Rob and Alice-AntiRob. The subsystem which is effectively two level is that of Alice, i.e., of the inertial observer, and hence the method allows to compute the LAI and LII for measurements made on the subsystem probed by the inertial observer. In other words, we are able to plot the LAI $J^\rightarrow(\rho_{AR})$ and $J^\rightarrow(\rho_{A\bar{R}})$, and the LII $\mathcal{D}^\rightarrow(\rho_{AR})$ and $\mathcal{D}^\rightarrow(\rho_{A\bar{R}})$. We plot, in Fig. (\ref{fig:all-correlations-rindler-M-I-squeeze}), all correlations - mutual information, LAI and LII - for the Alice-Rob bipartition, with measurements carried out by Alice as a function of the squeezing parameter. Remark that the classical and quantum correlation differ quantitatively, while showing a similar behavior. We plot the same, in Fig. (\ref{fig:all-correlations-rindler-M-II-squeeze}), for the Alice-AntiRob bipartition.

\begin{figure}[ht]
\includegraphics[width=\linewidth]{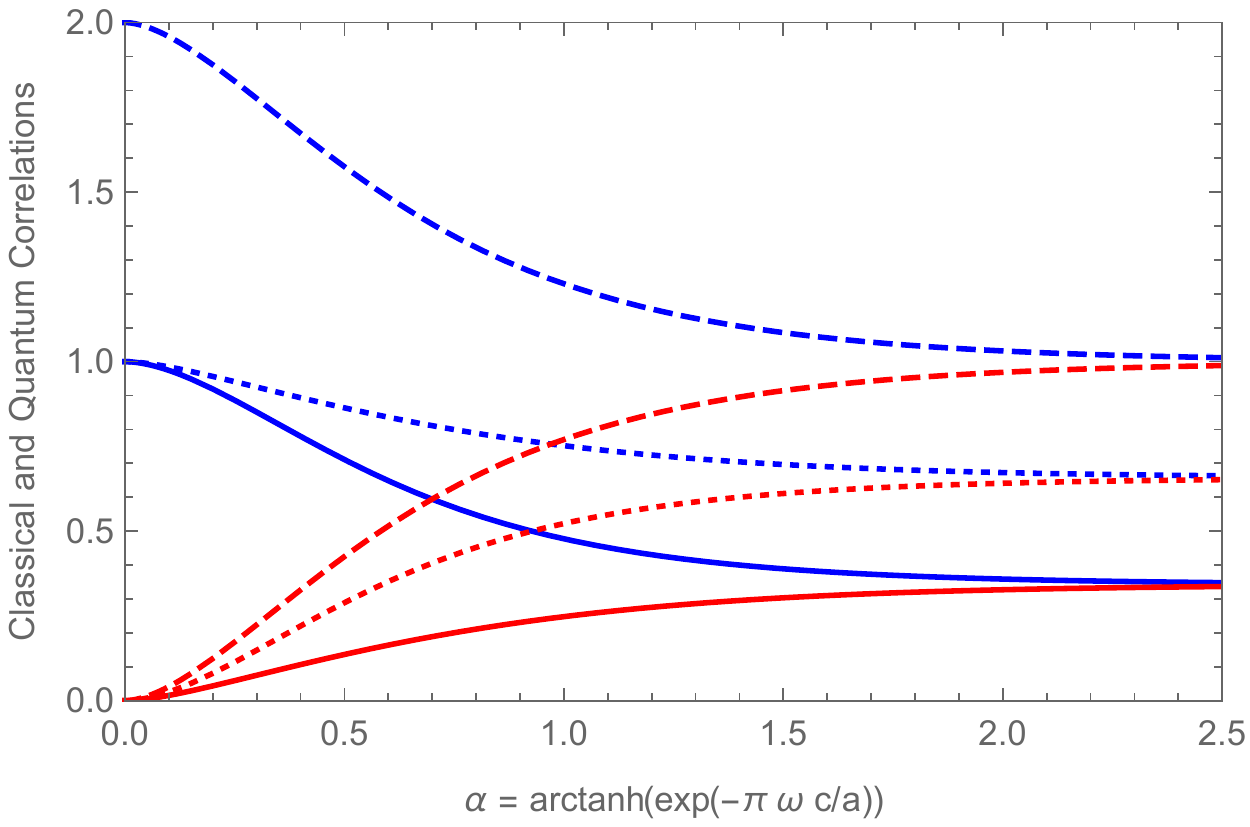}
\centering
\caption{States $\rho_{AR}$ and $\rho_{A\bar{R}}$ compared - The state $\rho_{AR}$ is depicted by the blue lines and $\rho_{A\bar{R}}$ by the red lines. Classical Correlations are the solid lines, quantum discord are the dotted lines and mutual information are dashed lines.}
\label{fig:all-correlations-rindler-compared-squeeze}
\end{figure}

Finally, to discuss the results, it is very instructive to plot all correlations measures (LAI, LII and mutual information) for the two bipartitions together. Doing so, using different colors for each bipartition we obtain the plot shown in Fig. (\ref{fig:all-correlations-rindler-compared-squeeze}).
There, the blue lines correspond to the bipartition between the inertial observer and the right Rindler observer, with the classical correlations and quantum discord characterizing respectively the locally accessible and locally inaccessible information for the inertial observer. The red lines are the plots for the bipartition among the inertial observer and the left Rindler observer and now the classical correlations and quantum discord characterize respectively the locally accessible and locally inaccessible information for the inertial observer. The case of zero acceleration and hence zero squeezing parameter is obviously the case in which we are considering just inertial observers. Hence we clearly see in the plot that when there is a non-zero acceleration, compared to the situation in which there is not, a trade-off of the correlations occur. 

\section{Entanglement of Formation}

Entanglement of formation is a measure of entanglement for mixed states that satisfies the basic requirements one would expect of an entanglement measure \cite{geometry-quantum-states}. If $\rho_{AB}$ is such a state, one defines the entanglement of formation to be
\begin{eqnarray}
        E_F(\rho_{AB})&=& \inf_{\{(|\psi_i\rangle,p_i)\}}\sum_i p_i S\bigg(\operatorname{Tr}_A(|\psi_i\rangle\langle \psi_i|)\bigg)\nonumber\\&=&\inf_{\{(|\psi_i\rangle,p_i)\}}\sum_i p_i S\bigg(\operatorname{Tr}_B(|\psi_i\rangle\langle \psi_i|)\bigg),
\end{eqnarray}
where the infimum is taken over the set of all ensembles of pure states that realize $\rho_{AB}$.

This measure has an operational interpretation that makes it valuable for applications. Still, in the present case, it is worth considering it because of its special connection to LAI and LII \cite{koashi-winter,marcos-cesar,marcos-cesar2}.

This connection lies in the relation that if $\rho_{ABC} = |\psi\rangle\langle \psi|$ is a tripartite pure state, then the entanglement of formation of the $AB$ subsystem is connected to the LAI of the $AC$ subsystem by means of the equation
\begin{equation}\label{eq:koashi-winter-relation}
    E_F(\rho_{AB})+J^\leftarrow(\rho_{AC})=S(\rho_A).
\end{equation}
By varying the subsystems one obtains other equations like that \cite{koashi-winter}. This relation has been employed in \cite{marcos-cesar} in order to obtain an important monogamy relation between entanglement of formation and quantum discord.

This relation has a very important impact on the interpretation of what entanglement of formation is quantifying. By rewriting the equation as \begin{equation}
   J^\leftarrow(\rho_{AC})=S(\rho_A)-E_F(\rho_{AB}),
\end{equation}
and recalling that $J^\leftarrow(\rho_{AC})$ is the information contained in the correlations between $A$ and $C$ locally accessible to $C$ by measurements, and recalling that $S(\rho_A)$ is the uncertainty in the state of $A$, we see that when $E_F(\rho_{AB}) = 0$, \textit{all the information} is locally accessible, and when $E_F(\rho_{AB}) > 0$ the locally accessible information decreases. In that setting, $E_F(\rho_{AB})$ signals a correlation redistribution by which an observer of $C$ alone looses access to information contained in the correlations of its state with that of the $A$ subsystem. This interpretation of $E_F(\rho_{AB})$ demands no "entanglement as a resource'' argument, and is available even if the subsystems $A$ and $B$ are separated by a causal horizon and the corresponding observers are forbidden classical communication.

Eq. (\ref{eq:koashi-winter-relation}) immediately implies that if $C$ is effectively two-level, in the sense of Eq. (\ref{eq:effectively-two-level}), then the method of \cite{datta}, which we have outlined  in the previous section, allows for the evaluation of $J^\leftarrow(\rho_{AC})$, and hence of $E_F(\rho_{AB})$.
\begin{figure}[ht]
    \centering
    \includegraphics[width=\linewidth]{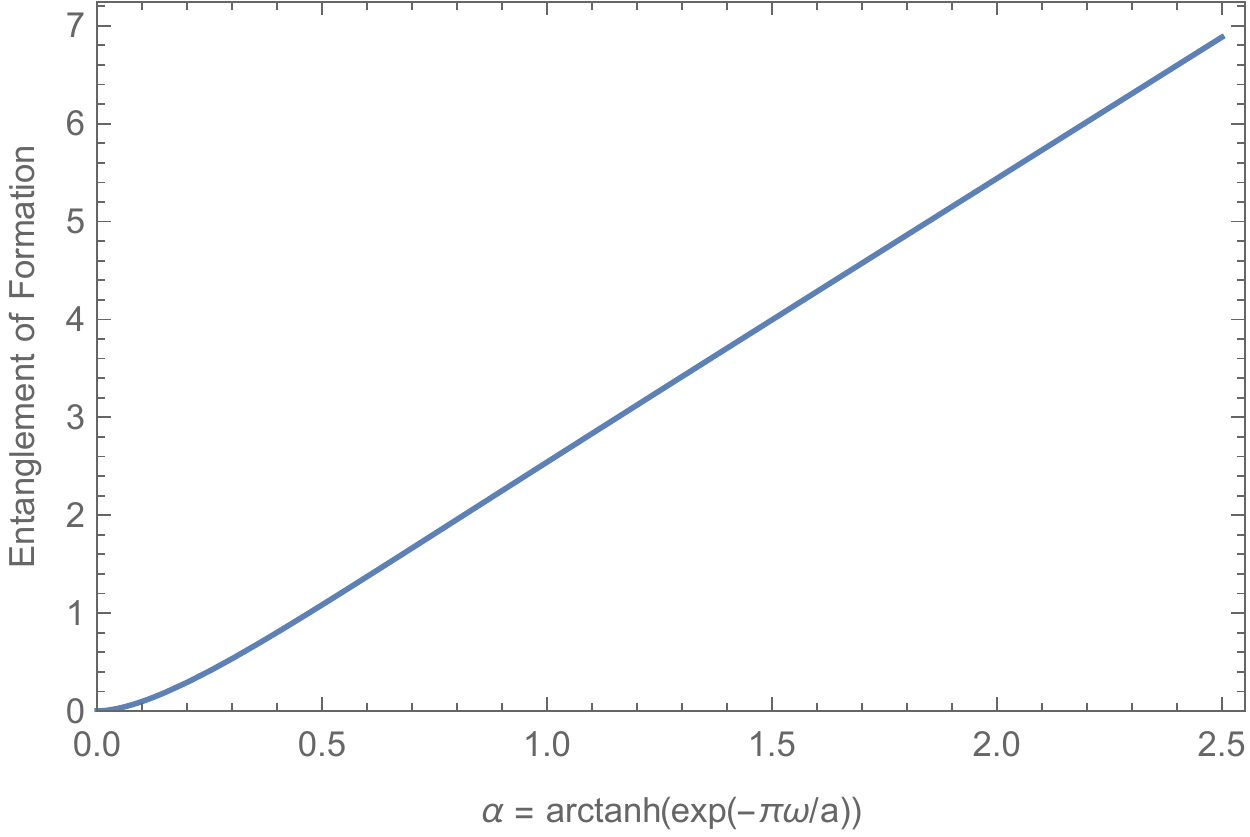}
    \caption{State $\rho_{R\bar{R}}$ - Entanglement of formation}
    \label{fig:entanglement-of-formation-rindler-i-ii}
\end{figure}

For the concrete problem we are considering, the only effectively two-level state is that of the subsystem probed by the inertial observer alone, which corresponds to the mode $u_i$ of the field. This allowed us the evaluation of $J^\rightarrow(\rho_{AR})$ and $J^\rightarrow(\rho_{A\bar{R}})$. Both LAIs, on the other hand, lead to \textit{the same} entanglement of formation, because the entanglement of formation is symmetric, i.e. $E_F(\rho_{R\bar{R}})=E_F(\rho_{\bar{R}R})$. The obtained entanglement of formation is shown in Fig. (\ref{fig:entanglement-of-formation-rindler-i-ii}) and we notice that it matches the overall behavior of the negativity computed in \cite{martin-martinez}.

We must stress that considering that the state $\rho_{R\bar{R}}$ is relatively complex (c.f. Eq. (\ref{eq:state-rob-antirob})) it is remarkable that we are able to compute the entanglement of formation, which is defined by a very difficult optimization, by just optimizing over two angles!

\section{Discussion and Conclusions}

We revisited the analysis of correlations of a two-mode state of a massless Klein-Gordon field which, for two inertial observers, Alice and Bob, is maximally entangled, when Bob is replaced by the Rindler observers Rob and AntiRob on respectively the right and left Rindler wedges. Our focus has been on informational quantities: the locally accessible and locally unaccessible informations, and the entanglement of formation directly connected to the previous two.

We built upon the method of \cite{datta} and evaluated both LAI and LII for both the Alice-Rob and Alice-AntiRob bipartitions and found a correlation redistribution associated to both quantities. Moreover, the ideas of \cite{koashi-winter,marcos-cesar,marcos-cesar2} allowed us to use these results to evaluate the entanglement of formation for the Rob-AntiRob bipartition. Given its relation to LAI and LII we are led to interpret it as a quantifier of the correlation redistribution. Our conclusion is that the causal horizon affecting the Rindler observers impart a correlation redistribution on the system when compared to the situation on which it is probed by two observer that neither perceive such causal horizon. Furthermore, the correlation redistribution imparted by the causal horizon seems to be quantified by the entanglement across the horizon. In that sense, even though such entanglement cannot be employed as a resource for any quantum computation task because the Rob-AntiRob bipartition is deprived of classical communication, the Entanglement of Formation is an important measure for quantification of the information content in accelerated frames in a consistent way. In fact, this approach seems highly significant for the investigation of Hawking radiation in Black Hole evaporation and the distribution of information content - a problem which is to be discussed elsewhere. 
\section*{Acknowledgements}
This study was financed in part by the Coordena\c c\~ao de Aperfei\c coamento de Pessoal de N\'\i vel Superior - Brasil (CAPES) - Finance Code 001 and by Conselho Nacional de Desenvolvimento Cient\'\i fico e Tecnol\'ogico (CNPq, process number 132437/2017-1). The authors acknowledge insightful discussions with A. Saa and G.E.A. Matsas, regarding the Unruh effect.

\bibliography{main}

\begin{thebibliography}{17}%
\makeatletter
\providecommand \@ifxundefined [1]{%
 \@ifx{#1\undefined}
}%
\providecommand \@ifnum [1]{%
 \ifnum #1\expandafter \@firstoftwo
 \else \expandafter \@secondoftwo
 \fi
}%
\providecommand \@ifx [1]{%
 \ifx #1\expandafter \@firstoftwo
 \else \expandafter \@secondoftwo
 \fi
}%
\providecommand \natexlab [1]{#1}%
\providecommand \enquote  [1]{``#1''}%
\providecommand \bibnamefont  [1]{#1}%
\providecommand \bibfnamefont [1]{#1}%
\providecommand \citenamefont [1]{#1}%
\providecommand \href@noop [0]{\@secondoftwo}%
\providecommand \href [0]{\begingroup \@sanitize@url \@href}%
\providecommand \@href[1]{\@@startlink{#1}\@@href}%
\providecommand \@@href[1]{\endgroup#1\@@endlink}%
\providecommand \@sanitize@url [0]{\catcode `\\12\catcode `\$12\catcode
  `\&12\catcode `\#12\catcode `\^12\catcode `\_12\catcode `\%12\relax}%
\providecommand \@@startlink[1]{}%
\providecommand \@@endlink[0]{}%
\providecommand \url  [0]{\begingroup\@sanitize@url \@url }%
\providecommand \@url [1]{\endgroup\@href {#1}{\urlprefix }}%
\providecommand \urlprefix  [0]{URL }%
\providecommand \Eprint [0]{\href }%
\providecommand \doibase [0]{http://dx.doi.org/}%
\providecommand \selectlanguage [0]{\@gobble}%
\providecommand \bibinfo  [0]{\@secondoftwo}%
\providecommand \bibfield  [0]{\@secondoftwo}%
\providecommand \translation [1]{[#1]}%
\providecommand \BibitemOpen [0]{}%
\providecommand \bibitemStop [0]{}%
\providecommand \bibitemNoStop [0]{.\EOS\space}%
\providecommand \EOS [0]{\spacefactor3000\relax}%
\providecommand \BibitemShut  [1]{\csname bibitem#1\endcsname}%
\let\auto@bib@innerbib\@empty
%</preamble>
\bibitem [{\citenamefont {Unruh}(1976)}]{unruh}%
  \BibitemOpen
  \bibfield  {author} {\bibinfo {author} {\bibfnamefont {W.~G.}\ \bibnamefont
  {Unruh}},\ }\href@noop {} {\bibfield  {journal} {\bibinfo  {journal}
  {Physical Review D}\ }\textbf {\bibinfo {volume} {14}},\ \bibinfo {pages}
  {870} (\bibinfo {year} {1976})}\BibitemShut {NoStop}%
\bibitem [{\citenamefont {Mart{\'\i}n-Mart{\'\i}nez}\ \emph
  {et~al.}(2010)\citenamefont {Mart{\'\i}n-Mart{\'\i}nez}, \citenamefont
  {Garay},\ and\ \citenamefont {Le{\'o}n}}]{martin-martinez}%
  \BibitemOpen
  \bibfield  {author} {\bibinfo {author} {\bibfnamefont {E.}~\bibnamefont
  {Mart{\'\i}n-Mart{\'\i}nez}}, \bibinfo {author} {\bibfnamefont {L.~J.}\
  \bibnamefont {Garay}}, \ and\ \bibinfo {author} {\bibfnamefont
  {J.}~\bibnamefont {Le{\'o}n}},\ }\href@noop {} {\bibfield  {journal}
  {\bibinfo  {journal} {Physical review D}\ }\textbf {\bibinfo {volume} {82}},\
  \bibinfo {pages} {064006} (\bibinfo {year} {2010})}\BibitemShut {NoStop}%
\bibitem [{\citenamefont {Datta}(2009)}]{datta}%
  \BibitemOpen
  \bibfield  {author} {\bibinfo {author} {\bibfnamefont {A.}~\bibnamefont
  {Datta}},\ }\href@noop {} {\bibfield  {journal} {\bibinfo  {journal}
  {Physical Review A}\ }\textbf {\bibinfo {volume} {80}},\ \bibinfo {pages}
  {052304} (\bibinfo {year} {2009})}\BibitemShut {NoStop}%
\bibitem [{\citenamefont {Koashi}\ and\ \citenamefont
  {Winter}(2004)}]{koashi-winter}%
  \BibitemOpen
  \bibfield  {author} {\bibinfo {author} {\bibfnamefont {M.}~\bibnamefont
  {Koashi}}\ and\ \bibinfo {author} {\bibfnamefont {A.}~\bibnamefont
  {Winter}},\ }\href {\doibase 10.1103/PhysRevA.69.022309} {\bibfield
  {journal} {\bibinfo  {journal} {Phys. Rev. A}\ }\textbf {\bibinfo {volume}
  {69}},\ \bibinfo {pages} {022309} (\bibinfo {year} {2004})}\BibitemShut
  {NoStop}%
\bibitem [{\citenamefont {Fanchini}\ \emph {et~al.}(2011)\citenamefont
  {Fanchini}, \citenamefont {Cornelio}, \citenamefont {de~Oliveira},\ and\
  \citenamefont {Caldeira}}]{marcos-cesar}%
  \BibitemOpen
  \bibfield  {author} {\bibinfo {author} {\bibfnamefont {F.~F.}\ \bibnamefont
  {Fanchini}}, \bibinfo {author} {\bibfnamefont {M.~F.}\ \bibnamefont
  {Cornelio}}, \bibinfo {author} {\bibfnamefont {M.~C.}\ \bibnamefont
  {de~Oliveira}}, \ and\ \bibinfo {author} {\bibfnamefont {A.~O.}\ \bibnamefont
  {Caldeira}},\ }\href@noop {} {\bibfield  {journal} {\bibinfo  {journal}
  {Physical Review A}\ }\textbf {\bibinfo {volume} {84}},\ \bibinfo {pages}
  {012313} (\bibinfo {year} {2011})}\BibitemShut {NoStop}%
\bibitem [{\citenamefont {Fanchini}\ \emph {et~al.}(2012)\citenamefont
  {Fanchini}, \citenamefont {Castelano}, \citenamefont {Cornelio},\ and\
  \citenamefont {de~Oliveira}}]{marcos-cesar2}%
  \BibitemOpen
  \bibfield  {author} {\bibinfo {author} {\bibfnamefont {F.~F.}\ \bibnamefont
  {Fanchini}}, \bibinfo {author} {\bibfnamefont {L.}~\bibnamefont {Castelano}},
  \bibinfo {author} {\bibfnamefont {M.~F.}\ \bibnamefont {Cornelio}}, \ and\
  \bibinfo {author} {\bibfnamefont {M.~C.}\ \bibnamefont {de~Oliveira}},\
  }\href {\doibase 10.1088/1367-2630/14/1/013027} {\bibfield  {journal}
  {\bibinfo  {journal} {New Journal of Physics}\ }\textbf {\bibinfo {volume}
  {14}},\ \bibinfo {pages} {013027} (\bibinfo {year} {2012})}\BibitemShut
  {NoStop}%
\bibitem [{Note1()}]{Note1}%
  \BibitemOpen
  \bibinfo {note} {In fact, the surfaces characterized by $t = \pm x $ are null
  surfaces respectively called past and future Rindler horizons, denoted
  $\protect \mathcal {H}^-$ and $\protect \mathcal {H}^+$, and they act as
  spacetime boundaries for the Rindler observers.}\BibitemShut {Stop}%
\bibitem [{\citenamefont {Fabbri}\ and\ \citenamefont
  {Navarro-Salas}(2005)}]{navarro-salas}%
  \BibitemOpen
  \bibfield  {author} {\bibinfo {author} {\bibfnamefont {A.}~\bibnamefont
  {Fabbri}}\ and\ \bibinfo {author} {\bibfnamefont {J.}~\bibnamefont
  {Navarro-Salas}},\ }\href@noop {} {\emph {\bibinfo {title} {{Modeling black
  hole evaporation}}}}\ (\bibinfo {year} {2005})\BibitemShut {NoStop}%
%%CITATION = INSPIRE-689791;%%
\bibitem [{\citenamefont {Birrell}\ and\ \citenamefont
  {Davies}(1984)}]{birrell-davies}%
  \BibitemOpen
  \bibfield  {author} {\bibinfo {author} {\bibfnamefont {N.~D.}\ \bibnamefont
  {Birrell}}\ and\ \bibinfo {author} {\bibfnamefont {P.~C.~W.}\ \bibnamefont
  {Davies}},\ }\href {\doibase 10.1017/CBO9780511622632} {\emph {\bibinfo
  {title} {{Quantum Fields in Curved Space}}}},\ Cambridge Monographs on
  Mathematical Physics\ (\bibinfo  {publisher} {Cambridge Univ. Press},\
  \bibinfo {address} {Cambridge, UK},\ \bibinfo {year} {1984})\BibitemShut
  {NoStop}%
%%CITATION = INSPIRE-181166;%%
\bibitem [{\citenamefont {Wald}(1994)}]{wald-qft}%
  \BibitemOpen
  \bibfield  {author} {\bibinfo {author} {\bibfnamefont {R.}~\bibnamefont
  {Wald}},\ }\href {https://books.google.com.br/books?id=Iud7eyDxT1AC} {\emph
  {\bibinfo {title} {Quantum Field Theory in Curved Spacetime and Black Hole
  Thermodynamics}}},\ Chicago Lectures in Physics\ (\bibinfo  {publisher}
  {University of Chicago Press},\ \bibinfo {year} {1994})\BibitemShut {NoStop}%
\bibitem [{\citenamefont {Wald}(1984)}]{wald}%
  \BibitemOpen
  \bibfield  {author} {\bibinfo {author} {\bibfnamefont {R.~M.}\ \bibnamefont
  {Wald}},\ }\href {\doibase 10.7208/chicago/9780226870373.001.0001} {\emph
  {\bibinfo {title} {{General Relativity}}}}\ (\bibinfo  {publisher} {Chicago
  Univ. Pr.},\ \bibinfo {address} {Chicago, USA},\ \bibinfo {year}
  {1984})\BibitemShut {NoStop}%
%%CITATION = INSPIRE-209356;%%
\bibitem [{\citenamefont {Duncan}(2012)}]{duncan}%
  \BibitemOpen
  \bibfield  {author} {\bibinfo {author} {\bibfnamefont {A.}~\bibnamefont
  {Duncan}},\ }\href@noop {} {\emph {\bibinfo {title} {The conceptual framework
  of quantum field theory}}}\ (\bibinfo  {publisher} {Oxford University
  Press},\ \bibinfo {year} {2012})\BibitemShut {NoStop}%
\bibitem [{\citenamefont {Weinberg}(1995)}]{weinberg}%
  \BibitemOpen
  \bibfield  {author} {\bibinfo {author} {\bibfnamefont {S.}~\bibnamefont
  {Weinberg}},\ }\href@noop {} {\emph {\bibinfo {title} {The quantum theory of
  fields. Vol. 1: Foundations}}}\ (\bibinfo  {publisher} {Cambridge University
  Press},\ \bibinfo {year} {1995})\BibitemShut {NoStop}%
\bibitem [{\citenamefont {Peskin}(2018)}]{peskin}%
  \BibitemOpen
  \bibfield  {author} {\bibinfo {author} {\bibfnamefont {M.~E.}\ \bibnamefont
  {Peskin}},\ }\href@noop {} {\emph {\bibinfo {title} {An introduction to
  quantum field theory}}}\ (\bibinfo  {publisher} {CRC Press},\ \bibinfo {year}
  {2018})\BibitemShut {NoStop}%
\bibitem [{\citenamefont {Schwartz}(2014)}]{schwartz}%
  \BibitemOpen
  \bibfield  {author} {\bibinfo {author} {\bibfnamefont {M.~D.}\ \bibnamefont
  {Schwartz}},\ }\href@noop {} {\emph {\bibinfo {title} {Quantum field theory
  and the standard model}}}\ (\bibinfo  {publisher} {Cambridge University
  Press},\ \bibinfo {year} {2014})\BibitemShut {NoStop}%
\bibitem [{\citenamefont {Carroll}\ \emph {et~al.}(2004)\citenamefont
  {Carroll}, \citenamefont {Carroll},\ and\ \citenamefont
  {Addison-Wesley}}]{carroll}%
  \BibitemOpen
  \bibfield  {author} {\bibinfo {author} {\bibfnamefont {S.}~\bibnamefont
  {Carroll}}, \bibinfo {author} {\bibfnamefont {S.}~\bibnamefont {Carroll}}, \
  and\ \bibinfo {author} {\bibnamefont {Addison-Wesley}},\ }\href
  {https://books.google.com.br/books?id=1SKFQgAACAAJ} {\emph {\bibinfo {title}
  {Spacetime and Geometry: An Introduction to General Relativity}}}\ (\bibinfo
  {publisher} {Addison Wesley},\ \bibinfo {year} {2004})\BibitemShut {NoStop}%
\bibitem [{\citenamefont {Bengtsson}\ and\ \citenamefont
  {{\.Z}yczkowski}(2017)}]{geometry-quantum-states}%
  \BibitemOpen
  \bibfield  {author} {\bibinfo {author} {\bibfnamefont {I.}~\bibnamefont
  {Bengtsson}}\ and\ \bibinfo {author} {\bibfnamefont {K.}~\bibnamefont
  {{\.Z}yczkowski}},\ }\href@noop {} {\emph {\bibinfo {title} {Geometry of
  quantum states: an introduction to quantum entanglement}}}\ (\bibinfo
  {publisher} {Cambridge university press},\ \bibinfo {year}
  {2017})\BibitemShut {NoStop}%
\end{thebibliography}%

\end{document}